\documentclass[runningheads]{llncs}
\usepackage{amsmath,graphicx}
\usepackage{float}
\graphicspath{ {./figs/} }
\usepackage[T1]{fontenc}

\usepackage{booktabs}
\usepackage{subcaption}

\begin{document}
\title{MedSegMamba: 3D CNN-Mamba Hybrid Architecture for Brain Segmentation}
\titlerunning{MedSegMamba: 3D CNN-Mamba for Brain Segmentation}
%
%
%
\author{Aaron Cao\inst{1} \and Zongyu Li\inst{2} \and Jordan Jomsky\inst{3} \and Andrew F. Laine\inst{2} \and Jia Guo\inst{2,4,5}\thanks{Correspondence: Jia Guo, jg3400@columbia.edu}}
\authorrunning{A. Cao et al.}
\institute{University of California, Santa Barbara, Santa Barbara, CA, USA \and Department of Biomedical Engineering, Columbia University, New York, NY, USA \and Department of Data Science, Columbia University, New York, NY, USA \and Department of Psychiatry, Columbia University, New York, NY, USA \and Mortimer B. Zuckerman Mind Brain Behavior Institute, Columbia University, New York, NY, USA}

\maketitle              
\begin{abstract}

Widely used traditional pipelines for subcortical brain segmentation are often inefficient and slow, particularly when processing large datasets. Furthermore, deep learning models face challenges due to the high resolution of MRI images and the large number of anatomical classes involved. To address these limitations, we developed a 3D patch-based hybrid CNN-Mamba model that leverages Mamba’s selective scan algorithm, thereby enhancing segmentation accuracy and efficiency for 3D inputs.
This retrospective study utilized 1784 T1-weighted MRI scans from a diverse, multi-site dataset of healthy individuals. The dataset was divided into training, validation, and testing sets with a 1076/345/363 split. The scans were obtained from 1.5T and 3T MRI machines.
Our model's performance was validated against several benchmarks, including other CNN-Mamba, CNN-Transformer, and pure CNN networks, using FreeSurfer-generated ground truths. We employed the Dice Similarity Coefficient (DSC), Volume Similarity (VS), and Average Symmetric Surface Distance (ASSD) as evaluation metrics. Statistical significance was determined using the Wilcoxon signed-rank test with a threshold of \textit{P} < 0.05.
The proposed model achieved the highest overall performance across all metrics (DSC 0.88383; VS 0.97076; ASSD 0.33604), significantly outperforming all non-Mamba-based models (\textit{P} < 0.001). While the model did not show significant improvement in DSC or VS compared to another Mamba-based model (\textit{P}-values of 0.114 and 0.425), it demonstrated a significant enhancement in ASSD (\textit{P} < 0.001) with approximately 20\% fewer parameters.
In conclusion, our proposed hybrid CNN-Mamba architecture offers an efficient and accurate approach for 3D subcortical brain segmentation, demonstrating potential advantages over existing methods. Code is available at: \url{https://github.com/aaroncao06/MedSegMamba}.

\end{abstract}
\section{Introduction}
\label{sec:intro}

Subcortical brain segmentation is a significant application in medical image processing, as it enables the extraction of quantitative structural information from MRI scans. These localized details can aid in detecting and monitoring morphological deficits in various neuropsychiatric conditions, including Schizophrenia \cite{gutman2022meta}, Major Depressive Disorder \cite{ho2022subcortical}, and Dementia \cite{van2023subcortical}. However, achieving accurate brain segmentation has remained a formidable challenge due to the intricate 3D structures within the brain, the large number of anatomical labels, and the substantial computational resources required to process MRI scans at full resolution.

While manual segmentation stands as the most trusted method, it is a tedious and difficult task, even for experienced clinicians. Automated tools like FreeSurfer \cite{fischl2002whole} have been developed to address these challenges, providing a widely accepted standard for subcortical segmentation. Despite its widespread use, FreeSurfer's traditional methods can be slow—requiring many hours to process a single scan—and are often sensitive to data quality issues, which limits their reliability when processing large, heterogeneous datasets.

The FastSurfer pipeline \cite{henschel2020fastsurfer} \cite{henschel2022fastsurfervinn} is recognized as one of the foremost deep learning-based alternatives to FreeSurfer, capable of executing whole-brain level segmentations. For a 2.5D approach, FastSurfer aggregates three 2D fully convolutional neural networks that utilize the classic encoder-decoder structure originating from the U-Net \cite{ronneberger2015u}. Even though FastSurfer has demonstrated superior performance compared to standard alternative models such as 3D U-Net \cite{cciccek20163d}, QuickNAT \cite{GUHAROY2019713}, and SDNet \cite{sdnet}, its reliance on 2D models inherently limits its ability to capture the full 3D spatial dependencies of the brain's anatomical structures.

On the other hand, 3D patch-based solutions are better suited to capture such geometries. Although full 3D volume deep learning models for segmenting many classes are currently not feasible due to data and memory constraints, a patch-based approach significantly reduces memory usage and generates more training samples per subject. Additionally, smaller patches allow the model to better capture local 3D information, leading to more accurate segmentation in complex anatomical structures.

Moreover, traditional pure CNN architectures like those used in FastSurfer can suffer from the local receptive fields in each layer, making them susceptible to overlooking the comprehensive global 3D context. The Vision Transformer (ViT) architecture \cite{dosovitskiy2020vit} addresses this limitation by employing a self-attention mechanism—originally developed for natural language processing \cite{vaswani2017attention}—that achieves state-of-the-art performance in image recognition without relying on convolutions. Each self-attention layer has a global receptive field, enabling the model to extract deeper long-range spatial dependencies. Hybrid CNN-Transformer architectures modeled after the U-Net have become popular for medical image segmentation tasks, showing improved generalization and performance \cite{chen2021transunet} \cite{hatamizadeh2022unetr}. Specifically for subcortical segmentation, TABSurfer \cite{cao2023tabsurfer}, a CNN-Transformer hybrid building off of TABS \cite{rao2022improving} and inspired by TransBTS \cite{wang2021transbts}, demonstrated the advantages of combining a hybrid architecture with a 3D patch-based approach.

However, the computational cost of Transformers scales quadratically with sequence length, which leads to substantial hardware memory requirements. This makes their application to dense prediction tasks challenging, particularly for large, high-resolution biomedical images. A shifted window-based self-attention approach \cite{liu2021swin} can mitigate this issue by reducing the computational burden, enabling the use of pure Transformer models. However, it also restricts the receptive fields in each layer, potentially leading to the loss of global relationships that ViT was designed to capture.

Recent advancements in state space models offer a promising alternative. Mamba \cite{mamba} introduces the selective scan mechanism (S6) with a hardware-aware algorithm to enhance both training and inference efficiency. This innovation allows the model to scale linearly with respect to sequence length while maintaining state-of-the-art performance across various long-sequence processing tasks. 

Since its introduction, Mamba has been adapted to various vision tasks, consistently demonstrating competitive results compared to both Transformer and CNN-based architectures, but with improved memory and parameter efficiency. For 2D image applications, Vision Mamba \cite{zhu2024vision} introduces the Vim block, which incorporates a bidirectional State-Space Model (SSM) combined with positional embeddings. Meanwhile, VMamba \cite{liu2024vmamba} introduced the Visual State-Space (VSS) block, centered around the 2D-Selective-Scan (SS2D) module. Because the visual components in images are not sequential, unlike the ordered nature of text and audio, the Vim block unravels the image in two ways (forward and backward) and the SS2D module unravels the image along four traversal paths, to mitigate any unidirectional bias in the final outputs. MambaAD \cite{he2024mambaad} demonstrates that processing additional traversal paths within each Mamba module can enhance performance in 2D anomaly detection tasks. However, in the realm of 3D segmentation, current methods do not fully exploit the richness of the 3D context. For example, U-Mamba \cite{U-Mamba} unravels the image in only one direction, and SegMamba's Tri-Oriented Mamba module \cite{xing2024segmamba} unfolds the image along just three directions.

In this paper, we introduce MedSegMamba, a hybrid CNN-Mamba architecture that fully leverages Mamba’s selective scan algorithm for 3D inputs, aiming to enhance subcortical brain segmentation.

\section{Materials and Methods}
\label{sec:methods}
This study did not require ethical approval as it utilized publicly available anonymous MRI scans previously acquired for studies approved by local institutional review boards, research ethics committees, or human investigation committees.

\subsection{Pipeline}
\label{ssec:pipeline}

\begin{figure*}[htb]
  \centering
  \includegraphics[width=.75\textwidth]{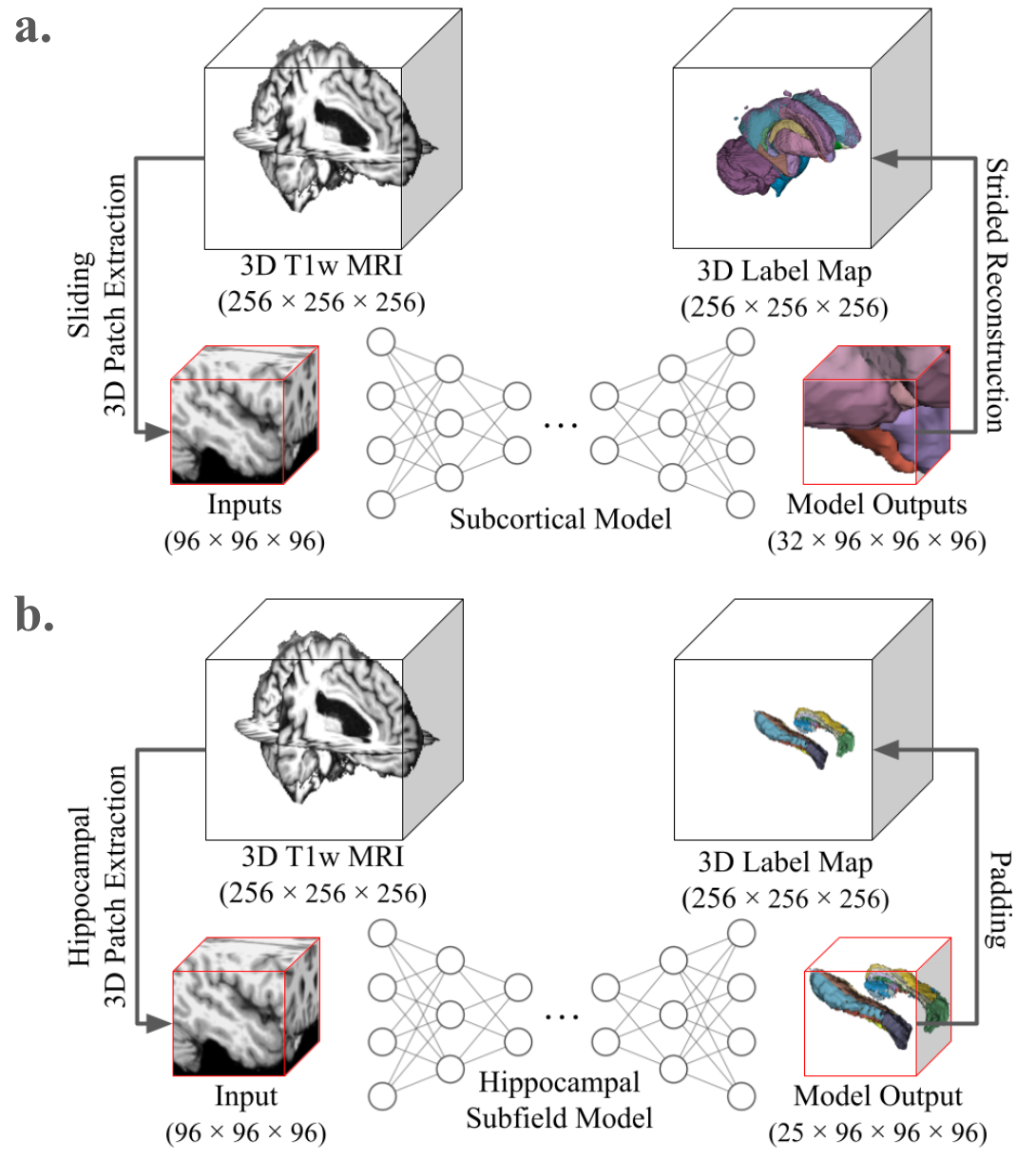}
  \caption{(a) The subcortical segmentation pipeline extracts 3D patches from the input scan, feeds them into the model, and reconstructs the output predicted label maps to generate the subcortical segmentation of the input scan. (b) The hippocampal subfield segmentation pipeline uses the subcortical segmentation label map to extract one patch centered on the hippocampus region. This patch is fed into the model and the output is padded to the original shape.}
  \label{fig:pipeline}
\end{figure*}

The subcortical segmentation pipeline employs a 3D patch-based approach with a hybrid CNN-Mamba model. Initially, the input scans are centered and conformed to LIA orientation, followed by intensity rescaling from 0 to 1, consistent with the steps in the FastSurfer pipeline. These input volumes with dimensions 256 x 256 x 256 are cropped and padded before patch extraction. Patches are then extracted with dimensions 96 x 96 x 96, with a step size of 16 voxels between consecutive patches. Each patch is sequentially fed into the model, and the output class probabilities are reconstructed to match the original input image dimensions. The predicted probabilities for each patch are combined through a voting mechanism to determine the class for each voxel, with the final values mapped to the corresponding FreeSurfer labels. As illustrated in Figure~\ref{fig:pipeline}a, this pipeline ensures that the model can segment an entire scan into 32 classes in under 90 seconds (all 31 subcortical structures covered by FastSurferVINN excluding cortical white matter).

The hippocampal subfield segmentation process is visualized in Figure~\ref{fig:pipeline}b. Using the left and right hippocampus classes in the subcortical segmentation, a bounding box with dimensions 96 x 96 x 96 is drawn around the hippocampus in the input scan, and this patch is fed to a separate hybrid CNN-Mamba model, which segments the hippocampal subfields. The output is then padded to the original input image dimensions. This segments the hippocampus into 25 classes according to FreeSurfer’s “FS60” level of hierarchy except with left and right labels separated.

\subsection{Model Architecture}
\label{ssec:model}

\begin{figure*}[htb]
  \centering
  \includegraphics[width=1\textwidth]{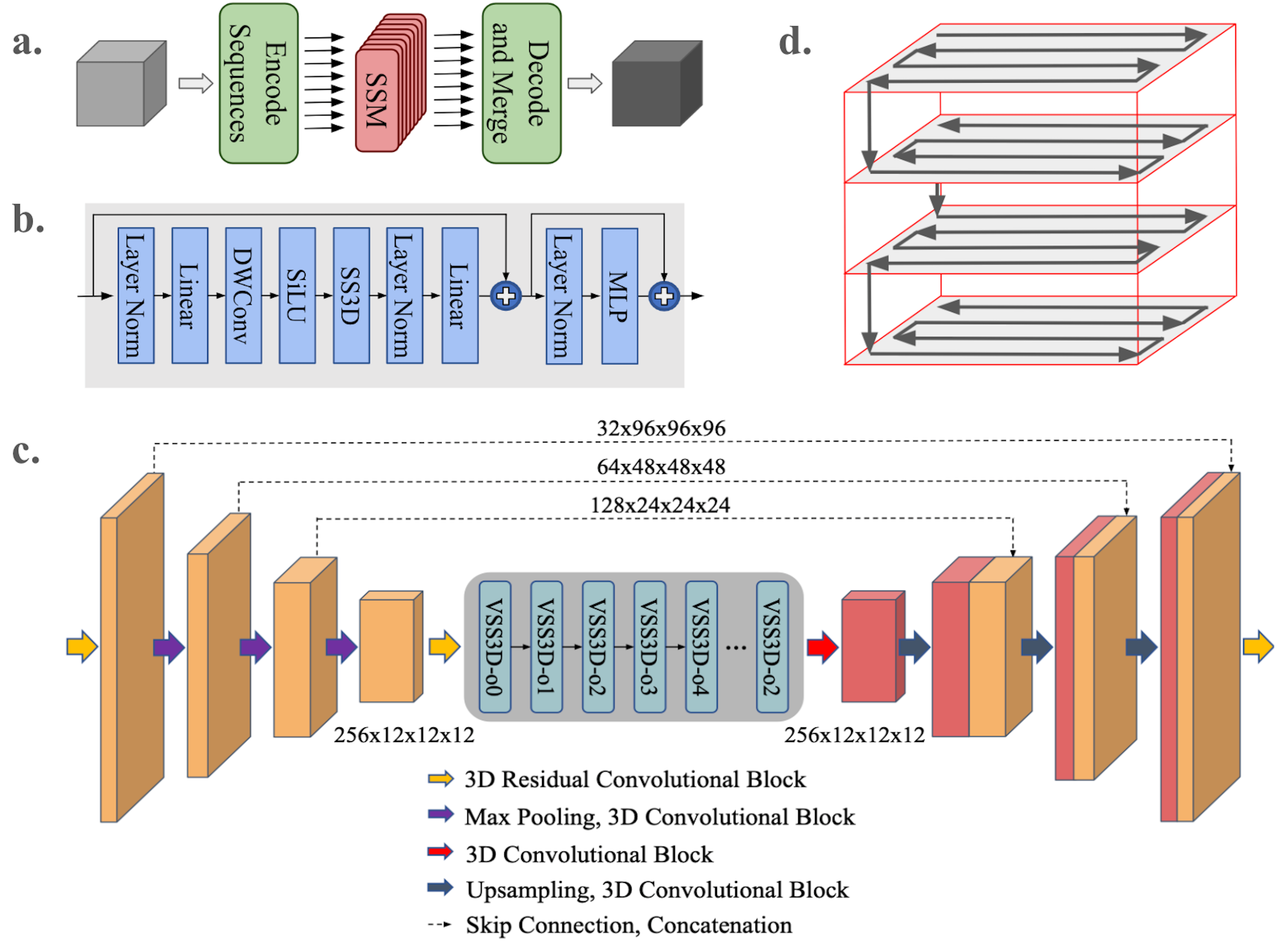}
  \caption{(a) The SS3D module encodes 8 sequences from the input volume, processes each with an independent S6 block, and then merges the outputs. (b) VSS3D block layout, consisting of SS3D and MLP residual modules. (c) The MedSegMamba model architecture has an encoder-decoder structure with a hybrid of CNN and Mamba-based blocks. Only 6 VSS3D layers are shown here, but the model contains 9 in total. The last Conv1x1 and Softmax layers are also not shown. (d) The SS3D modules unravel the sequences along this continuous 3D scanning pattern.
}
  \label{fig:architecture}
\end{figure*}

\subsubsection{3D Selective Scan (SS3D)}
\label{sssec:ss3d}

The selected 3D scanning pattern, depicted in Figure~\ref{fig:architecture}d, can unravel the volume along 48 unique traversal paths. Our core SS3D module, shown in Figure~\ref{fig:architecture}a, is designed to fully exploit each possible unfolded sequence, with each module processing one of six possible groups of eight sequences.  

First, the input volume's axes are transposed in one of the six possible ways (labeled o0 to o5), determined by the assigned orientation index of the SS3D module. The eight sequences are then extracted from the volume by rotating the volume in three different ways, unfolding it along the scanning pattern, and reversing the sequences. These eight sequences are processed independently and in parallel by the S6 blocks, with their outputs reordered and merged to reconstruct the final output volume. Each of the 48 methods of unraveling the volume is intended to train independent S6 blocks to capture the unique geometries best represented by that specific sequence order.

For the S6 blocks, we increase the SSM state dimension from the standard 16 to 64. Although this adjustment slightly reduces processing speed, it enables the model to extract more detailed information about complex structures with minimal impact on parameters and memory usage. The SSM feature expansion factor is set to 1 to keep the dimensionality constant, as the CNN encoder before the bottleneck has already expanded the feature dimension to 1024.

\subsubsection{VSS3D block}
\label{sssec:vss3d}

Our 3D Visual State Space (VSS3D) block resembles a traditional Transformer block, with self-attention replaced by our SS3D module, inspired by VMamba. It consists of two residual modules. The first module includes a sequence of layer normalization, linear projection, depth-wise convolution, SiLU activation, and SS3D, followed by another layer normalization and linear projection. The second residual module consists of a layer normalization followed by a multilayer perceptron (MLP). This is shown in Figure~\ref{fig:architecture}b.

\subsubsection{Overall Architecture}
\label{sssec:botarchitecture}

The overall model architecture features a 3D CNN encoder and decoder with skip connections and a series of VSS3D blocks as the bottleneck. Passing through the encoder, four layers of residual blocks and 3 max pooling operations downsample the input patch for an encoded feature tensor. A subsequent convolution brings the channel size to 1024, before the tensor is fed into the bottleneck. The bottleneck comprises a sequence of nine VSS3D blocks, each assigned an orientation index for its inner SS3D module. The output of the bottleneck is normalized and then passed to the decoder, which reconstructs the image to its original input dimensions. Finally, a convolution operation and a Softmax activation function are applied to generate either a 25 or 32 channel output, where each channel represents the probability of an individual class. The only difference between the subcortical and hippocampal subfield segmentation models is in the channel sizes of these last two layers. Each residual block within the encoder and decoder layers consists of a residual connection and two sequences of 3D Convolution, Group Normalization, and Rectified Linear Unit (ReLU). This is visualized in Figure~\ref{fig:architecture}c.

\subsection{Data}
\label{ssec:data}

For the subcortical segmentation dataset, a total of 1784 T1-weighted (T1w) MRI scans were selected from a large-scale heterogeneous dataset representing a uniformly healthy population, compiled from multiple publicly available sources \cite{feng2020estimating}. All scans were acquired at a resolution of 1mm x 1mm x 1mm. Selected subjects came from the Australian Imaging Biomarkers and Lifestyle Study of Ageing (AIBL) \cite{ellis2009australian}, Frontotemporal Lobar Degeneration Neuroimaging Initiative (NIFD) database (http://4rtniftldni.ini.usc.edu), Information eXtraction from Images (IXI) (http://brain-development.org/ixi-dataset), Open Access Series of Imaging Studies-1 (OASIS-1) \cite{marcus2007open}, Open Access Series of Imaging Studies-2 (OASIS-2) \cite{marcus2010open}, Southwest University Adult life-span Dataset (SALD) \cite{wei2018structural}, Southwest University Longitudinal Imaging Multimodal Brain Data Repository (SLIM) \cite{liu2017longitudinal}, Parkinson’s Progression Markers Initiative (PPMI) \cite{marek2011parkinson}, SchizConnect (SchizConnect) \cite{wang2016schizconnect}, and Consortium for Reliability and Reproducibility (CoRR) \cite{zuo2014open}. 

For the hippocampal subfield segmentation dataset, 1221 1mm x 1mm x 1mm T1w scans were selected from the same large-scale multi-site dataset \cite{feng2020estimating}. These subjects came from the Alzheimer’s Disease Neuroimaging Initiative (ADNI) database (adni.loni.usc.edu) and Brain Genomics Superstruct Project (BGSP) \cite{holmes2015brain}, in addition to AIBL, IXI, NIFD, and OAS1.

\begin{figure}[htb]
\centering
  \includegraphics[width=.85\textwidth]{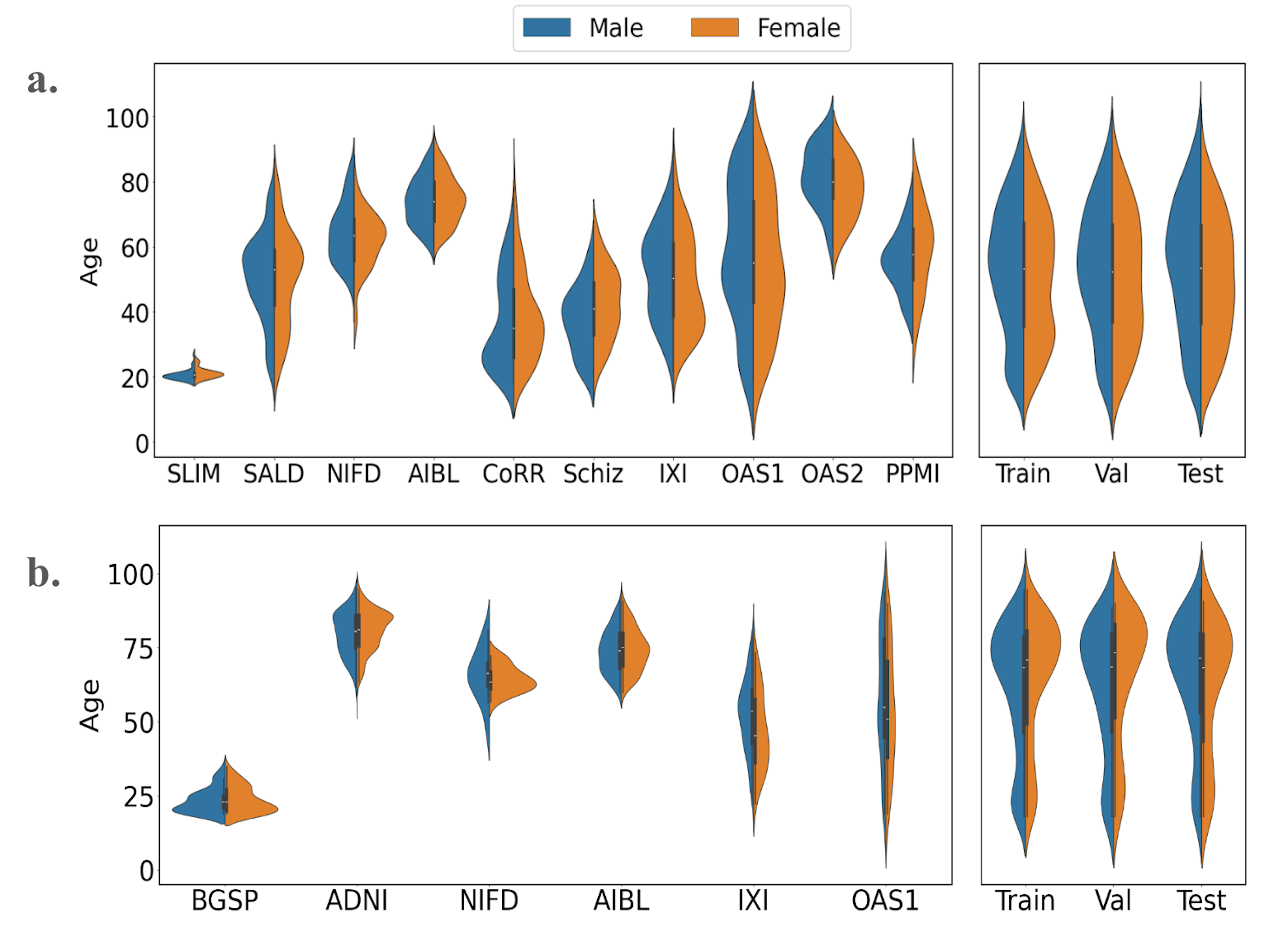}
  \caption{(a) Age and Gender Distributions for the Subcortical Segmentation Dataset (b) Age and Gender Distributions for the Hippocampal Subfield Segmentation Dataset.}
  \label{fig:data}
\end{figure}

Both datasets were partitioned into training, validation, and test sets with a roughly 3:1:1 ratio. Specifically, the subcortical dataset had a 1076/345/363 split, and the hippocampus dataset had a 719/254/248 split. A roughly balanced age and gender distribution was achieved across the datasets, as shown in Figure~\ref{fig:data}. 

Ground truth segmentations were generated using FreeSurfer, and the input T1w scans were preprocessed with skull-stripping and intensity normalization.

\subsection{Model Training}
\label{ssec:training}
The described model was trained and evaluated on a 24 GB NVIDIA Quadro 6000 GPU. We employed the AdamW optimizer in conjunction with a Cosine Annealing Warm Restarts learning rate scheduler \cite{loshchilov2016sgdr}. The loss function used was a combination of Dice Loss and Weighted Cross Entropy during the first epoch, followed by Dice Loss alone in subsequent epochs. Subcortical segmentation training was conducted for 15 epochs on 27 patches per scan (three steps per dimension), utilizing gradient accumulation to simulate each image as a single batch, accounting for the variability in class presence across patches. Hippocampal subfield segmentation training followed the same procedure for 63 epochs except with a batch size of 1. A dropout rate of 0.1 and a drop path rate of 0.3 were applied within the bottleneck.

\subsection{Model Evaluation and Statistical Test}
\label{ssec:eval}

To evaluate the efficacy of our novel SS3D module over other 3D mamba-based vision modules for latent space learning, we also trained a model identical to our MedSegMamba architecture, except substituting the first residual module in each VSS3D block with SegMamba's Tri-oriented Mamba module. The resulting block is identical to SegMamba's Tri-oriented Spatial Mamba block except without the Gated Spatial Convolution module, in order to focus specifically on evaluating the core Mamba modules. This model, referred to as SegMambaBot, was trained following the same procedures as MedSegMamba.

For subcortical segmentation, MedSegMamba was also evaluated against a pretrained FastSurferVINN (obtained from the FastSurfer GitHub repository) and a retrained TABSurfer. Both SegMambaBot and TABSurfer were trained using 3D patch-based approaches with input patch sizes identical to those used in MedSegMamba.

For hippocampal subfield segmentation, MedSegMamba was evaluated against SegMambaBot and TABSurfer, with each model being identical to their subcortical segmentation counterparts except for the final layers generating a 25 instead of 32 channel output. Each model was trained following the same procedure. FastSurfer was not evaluated for this task because only one patch containing the hippocampus needs to be processed, which is small enough for a 3D model to handle in one pass. Aggregating multiple 2D slice-based networks to process this smaller region would be slower and unnecessary. 

The evaluation metrics included the Dice Similarity Coefficient (DSC), Volume Similarity (VS) \cite{Taha2015}, and Average Symmetric Surface Distance (ASSD) \cite{assd}, assessing both the overall similarity of the segmentations and the contour quality relative to the ground truth. The Wilcoxon signed-rank test was employed to evaluate the significance of improvements in MedSegMamba’s performance over the other models for each metric. Statistical significance was set to \textit{P} < 0.05.

\section{Results}
\label{sec:results}

\subsection{Subcortical Segmentation}
\label{ssec:subcorticalresults}

\begin{figure*}
  \centering
  \includegraphics[width=.95\textwidth]{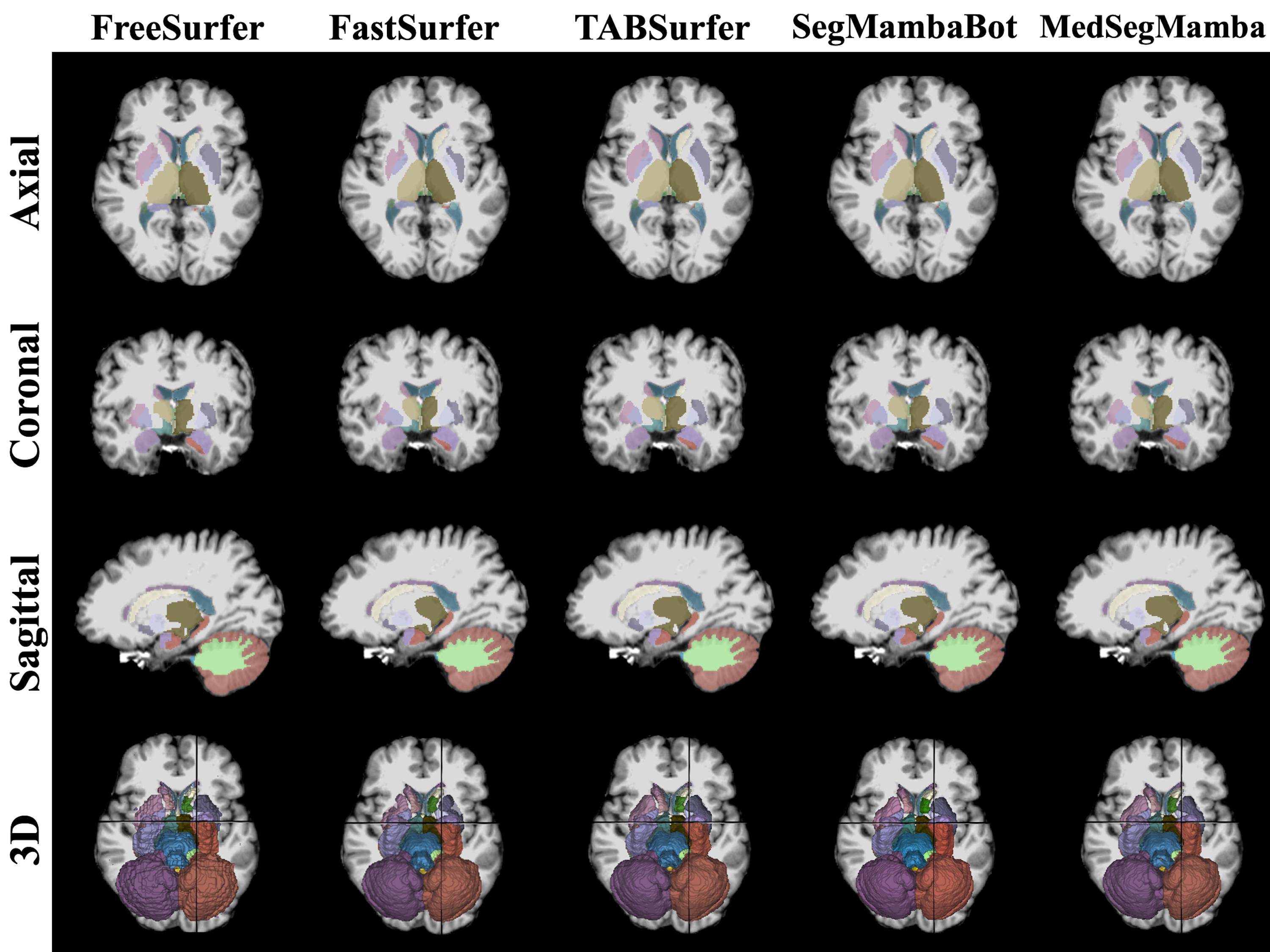}
  \caption{2D slices of a sample’s subcortical segmentation by each method.}
  \label{fig:2d_aseg}
\end{figure*}

The average metrics used for evaluating MedSegMamba, SegMambaBot, TABSurfer, and FastSurferVINN against the FreeSurfer-generated subcortical segmentation ground truths are displayed in Table~\ref{table:results}. MedSegMamba consistently achieved high scores in Dice Similarity Coefficient (DSC), Volume Similarity (VS), and Average Symmetric Surface Distance (ASSD) across all datasets, demonstrating superior overall performance. MedSegMamba’s improvement was statistically significant across all metrics, except for the DSC and VS metrics when compared to SegMamba (\textit{P}-values of 0.114 and 0.425, respectively). 

All three 3D patch-based methods also significantly outperformed the pretrained 2.5D FastSurfer benchmark. Furthermore, both Mamba-based models significantly outperformed TABSurfer, their transformer-based counterpart. 

\begin{figure*}
  \centering
  \includegraphics[width=1\textwidth]{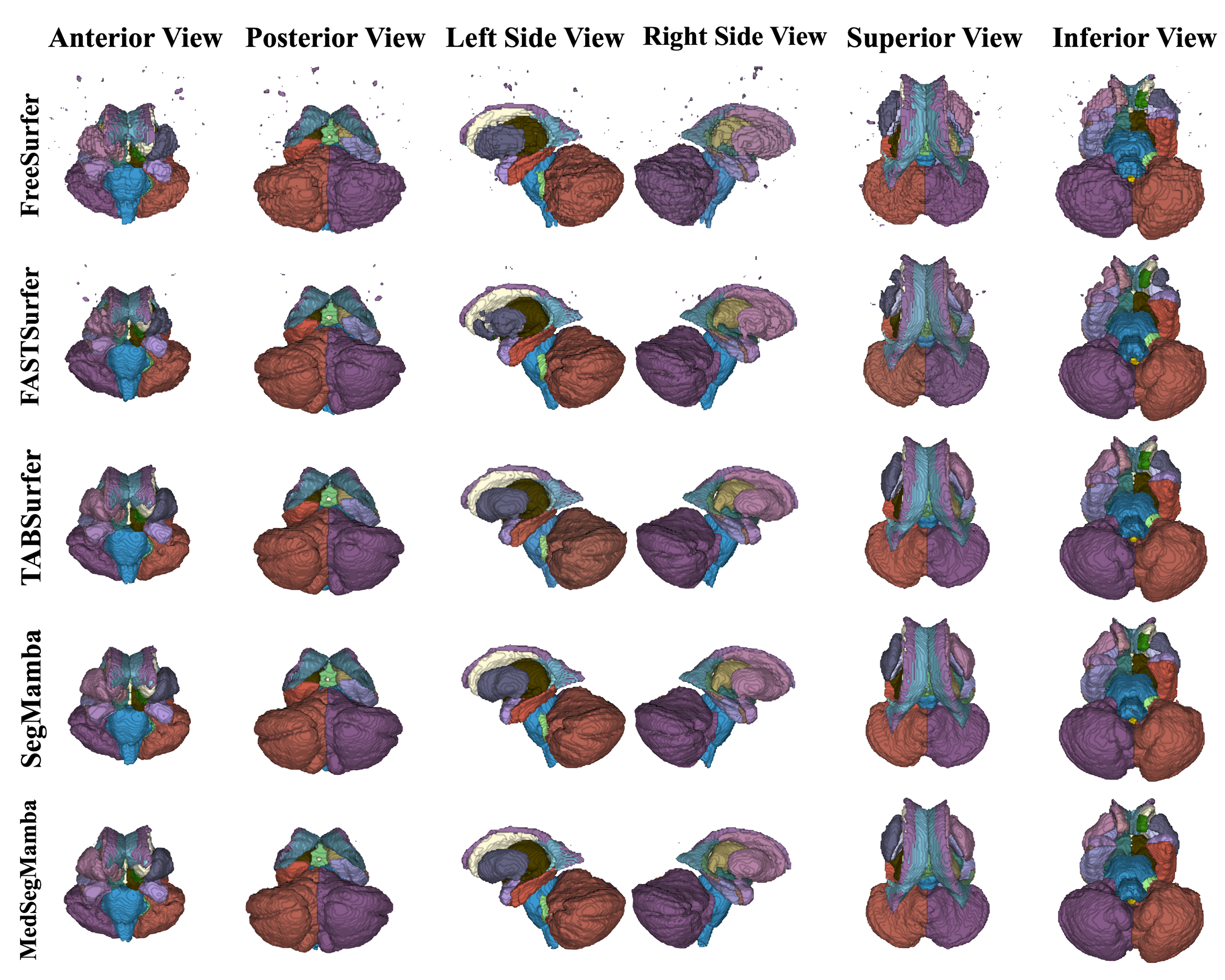}
  \caption{3D renderings of a sample’s subcortical segmentation by each method.}
  \label{fig:3d_aseg}
\end{figure*}

\begin{figure*}
  \centering
  \includegraphics[width=1\textwidth]{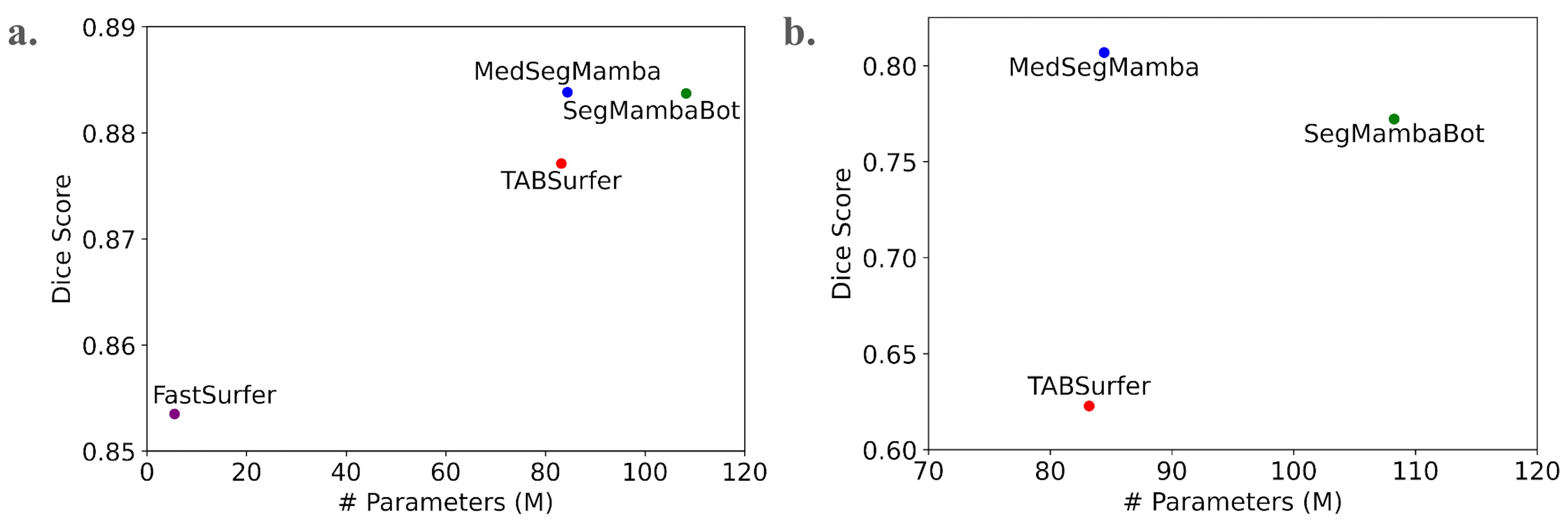}
  \caption{(a) Plot of millions of parameters versus subcortical segmentation dice score for each model. (b) Plot of millions of parameters versus hippocampal subfield segmentation dice score for each model.}
  \label{fig:parametersvsdice}
\end{figure*}

\addtolength{\tabcolsep}{1pt}  
\begin{table}[H]
\centering 
\begin{tabular}{lllll}
\toprule
Dataset & Model & DSC ↑ & VS ↑ & ASSD ↓ \\
\midrule
AIBL & \textbf{MedSegMamba} & \textbf{0.89593±0.009} & \textbf{0.97406±0.006} & 0.29069±0.042 \\
 & SegMambaBot & 0.89580±0.009 & 0.97396±0.006 & \textbf{0.29038±0.040} \\
 & TABSurfer & 0.89094±0.010 & 0.97082±0.007 & 0.30257±0.044 \\
 & FastSurfer & 0.87856±0.015 & 0.96485±0.009 & 0.33453±0.059 \\
\hline
CoRR & \textbf{MedSegMamba} & \textbf{0.88635±0.020} & \textbf{0.97123±0.008} & \textbf{0.31922±0.078} \\
 & SegMambaBot & 0.88627±0.020 & 0.97040±0.008 & 0.32062±0.077 \\
 & TABSurfer & 0.87949±0.020 & 0.96765±0.009 & 0.33669±0.074 \\
 & FastSurfer & 0.86607±0.027 & 0.95989±0.016 & 0.38024±0.104 \\
\hline
IXI & \textbf{MedSegMamba} & \textbf{0.86773±0.023} & \textbf{0.96485±0.010} & \textbf{0.41766±0.102} \\
 & SegMambaBot & 0.86729±0.023 & 0.96460±0.010 & 0.42425±0.106 \\
 & TABSurfer & 0.85877±0.027 & 0.96023±0.011 & 0.43849±0.108 \\
 & FastSurfer & 0.81217±0.034 & 0.93340±0.018 & 0.61388±0.140 \\
\hline
NIFD & \textbf{MedSegMamba} & \textbf{0.90060±0.007} & 0.97734±0.006 & \textbf{0.26797±0.030} \\
 & SegMambaBot & 0.90026±0.008 & \textbf{0.97745±0.005} & 0.27077±0.032 \\
 & TABSurfer & 0.89260±0.008 & 0.97361±0.005 & 0.29188±0.035 \\
 & FastSurfer & 0.88791±0.008 & 0.97240±0.005 & 0.30494±0.030 \\
\hline
OAS1 & MedSegMamba & 0.88954±0.012 & 0.97351±0.006 & 0.30856±0.050 \\
 & \textbf{SegMambaBot} & \textbf{0.88977±0.012} & \textbf{0.97387±0.007} & \textbf{0.30839±0.050} \\
 & TABSurfer & 0.88305±0.011 & 0.96808±0.005 & 0.32129±0.041 \\
 & FastSurfer & 0.87504±0.010 & 0.96240±0.005 & 0.34141±0.047 \\
\hline
OAS2 & \textbf{MedSegMamba} & \textbf{0.88680±0.013} & 0.97017±0.009 & \textbf{0.31079±0.047} \\
 & SegMambaBot & 0.88670±0.012 & \textbf{0.97048±0.008} & 0.31382±0.047 \\
 & TABSurfer & 0.88247±0.013 & 0.96611±0.009 & 0.32258±0.047 \\
 & FastSurfer & 0.87984±0.013 & 0.96454±0.008 & 0.32390±0.049 \\
\hline
PPMI & \textbf{MedSegMamba} & \textbf{0.89373±0.007} & 0.97345±0.005 & \textbf{0.29935±0.037} \\
 & SegMambaBot & 0.89333±0.007 & \textbf{0.97422±0.005} & 0.30096±0.036 \\
 & TABSurfer & 0.88941±0.008 & 0.97158±0.006 & 0.30483±0.032 \\
 & FastSurfer & 0.87892±0.008 & 0.96673±0.006 & 0.32847±0.033 \\
\hline
SALD & MedSegMamba & 0.87785±0.028 & 0.96998±0.009 & 0.35457±0.094 \\
 & \textbf{SegMambaBot} & \textbf{0.87795±0.028} & \textbf{0.97015±0.009} & \textbf{0.35236±0.093} \\
 & TABSurfer & 0.87046±0.027 & 0.96560±0.009 & 0.37473±0.092 \\
 & FastSurfer & 0.84189±0.021 & 0.94426±0.014 & 0.48194±0.089 \\
\hline
Schiz & \textbf{MedSegMamba} & 0.88204±0.011 & \textbf{0.97118±0.007} & \textbf{0.33188±0.044} \\
 & SegMambaBot & \textbf{0.88205±0.011} & 0.97082±0.008 & 0.33257±0.042 \\
 & TABSurfer & 0.87613±0.011 & 0.96747±0.007 & 0.34891±0.053 \\
 & FastSurfer & 0.83832±0.022 & 0.94248±0.015 & 0.48469±0.094 \\
\hline
SLIM & MedSegMamba & 0.88692±0.006 & 0.97101±0.006 & \textbf{0.30611±0.024} \\
 & \textbf{SegMambaBot} & \textbf{0.88740±0.007} & \textbf{0.97169±0.006} & 0.30679±0.025 \\
 & TABSurfer & 0.88254±0.006 & 0.96972±0.005 & 0.31617±0.022 \\
 & FastSurfer & 0.85483±0.012 & 0.94886±0.008 & 0.42471±0.051 \\
\hline
Overall & \textbf{MedSegMamba} & \textbf{0.88383±0.021} & \textbf{0.97076±0.009} & \textbf{0.33604±0.086} \\
 & SegMambaBot & 0.88372±0.021 & 0.97068±0.009 & 0.33761±0.087 \\
 & TABSurfer & 0.87711±0.022 & 0.96684±0.009 & 0.35264±0.088 \\
 & FastSurfer & 0.85350±0.035 & 0.95229±0.018 & 0.43554±0.143 \\
\bottomrule
\end{tabular}
\caption{Comparing MedSegMamba, SegMambaBot, TABSurfer, and FastsurferVINN metrics across datasets. Bold text indicates superior performance. ↑ indicates that higher numbers correspond to better performance and ↓ indicates that lower numbers correspond to better performance.}
    \label{table:results}  
\end{table}

Views of 3D renderings and 2D slices of segmentations produced by each method on a sample subject are shown in Figure~\ref{fig:3d_aseg} and Figure~\ref{fig:2d_aseg}. Although FreeSurfer’s atlas-based method yielded the noisiest segmentation, all deep learning methods produced smoother contours. FastSurfer’s 2.5D method retained the most noise from the FreeSurfer output, whereas the 3D patch-based methods achieved similarly smooth segmentations, differing only in minor areas for this sample. However, SegMamba exhibited slight under-segmentation in certain regions, and TABSurfer demonstrated less consistency across different scans.

While both Mamba-based models showed competitive performance, MedSegMamba demonstrated slight improvements over SegMambaBot and has approximately 20\% fewer parameters while requiring similar GPU memory during inference (1.988 GiB for MedSegMamba vs. 1.932 GiB for SegMambaBot). 

As shown in Figure~\ref{fig:parametersvsdice}a, MedSegMamba demonstrates superior performance while remaining parameter efficient compared to the other 3D patch-based methods. While FastSurfer's fully convolutional architecture requires very few parameters, each of its three sub-models requires more than double the GPU memory during inference (just under 5 GiB) compared to the 3D models, all of which operate with under 2 GiB of GPU memory.

\subsection{Hippocampal Subfield Segmentation}
\label{ssec:hippocampalresults}

\begin{figure*}
  \centering
  \includegraphics[width=1\textwidth]{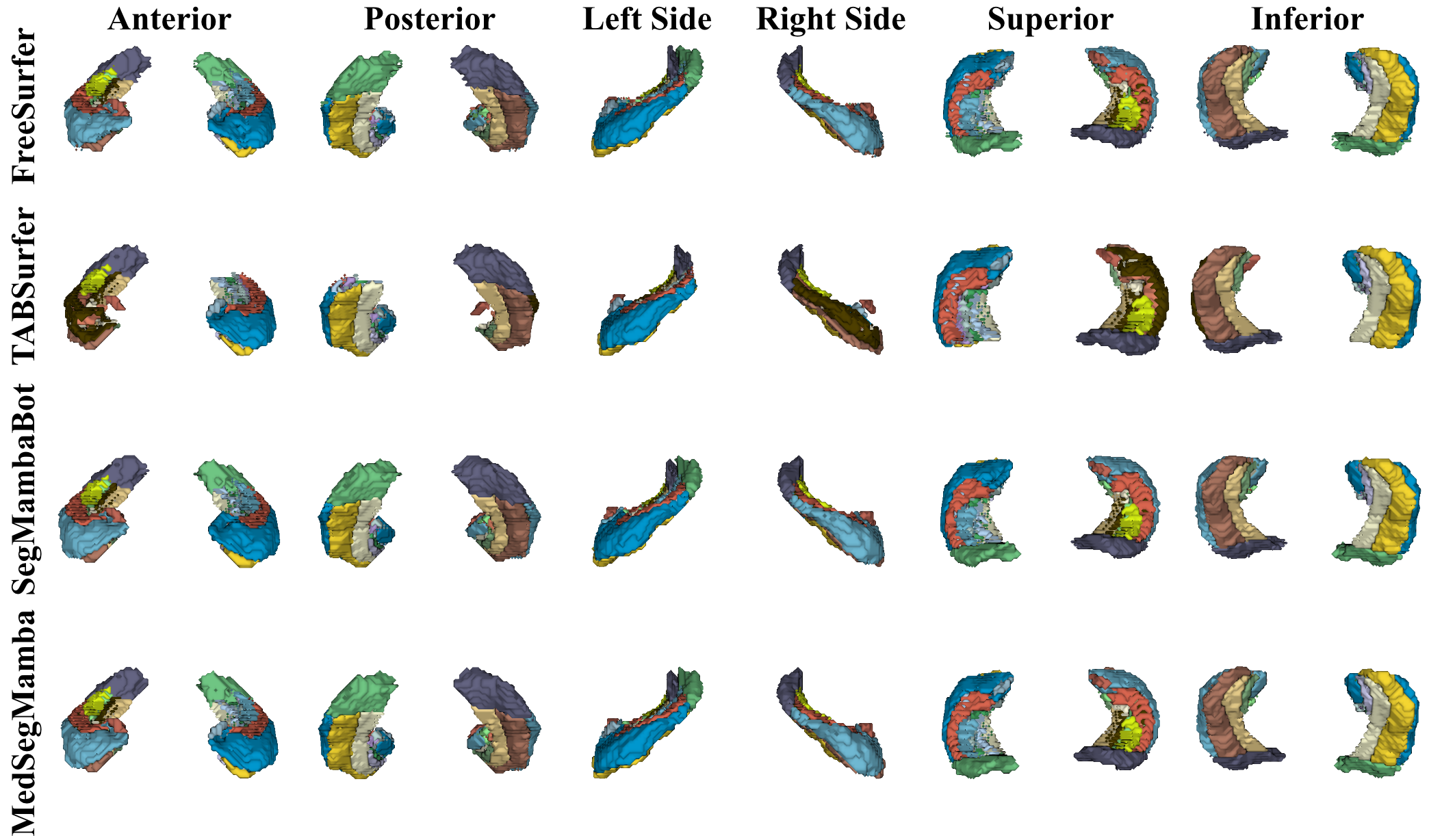}
  \caption{3D renderings of a sample’s hippocampal subfield segmentation by each method.}
  \label{fig:3d_hsf}
\end{figure*}

\begin{figure*}
  \centering
  \includegraphics[width=.95\textwidth]{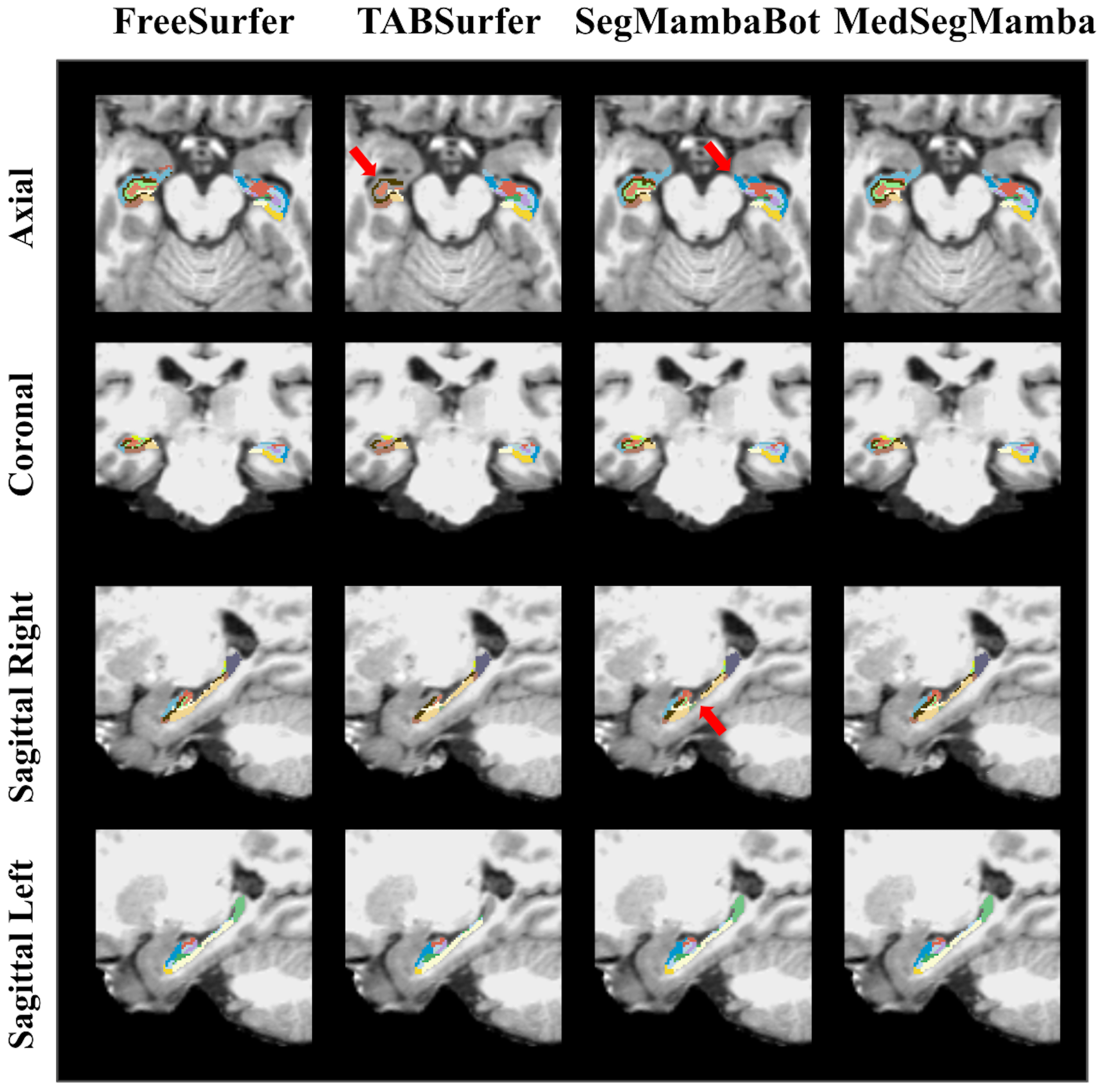}
  \caption{2D slices of a sample’s hippocampal subfield segmentation by each method are shown. Large regions in the right hippocampus are missing from TABSurfer’s segmentation, as indicated by the red arrow in the axial view. In SegMambaBot’s segmentation, the left hippocampal amygdala transition area is missing, as shown in the axial view, and the right presubiculum is slightly undersegmented, as shown in the sagittal right view.}
  \label{fig:2d_hsf}
\end{figure*}

\begin{table}[H]
\centering 
\begin{tabular}{lllll}
\toprule
Dataset & Model & DSC ↑ & VS ↑ & ASSD* ↓ \\
\midrule
ADNI & \textbf{MedSegMamba} & \textbf{0.80415±0.047} & \textbf{0.96207±0.014} & \textbf{0.22819±0.068} \\
 & SegMambaBot & 0.76976±0.045 & 0.92309±0.014 & --- \\
 & TABSurfer & 0.61932±0.037 & 0.75894±0.011 & --- \\
\hline
AIBL & \textbf{MedSegMamba} & \textbf{0.80935±0.037} & \textbf{0.96511±0.012} & \textbf{0.21981±0.055} \\
 & SegMambaBot & 0.77561±0.032 & 0.92525±0.010 & --- \\
 & TABSurfer & 0.62564±0.031 & 0.76095±0.009 & --- \\
\hline
BGSP & \textbf{MedSegMamba} & \textbf{0.81948±0.023} & \textbf{0.96375±0.010} & \textbf{0.21061±0.037} \\
 & SegMambaBot & 0.78562±0.018 & 0.92611±0.008 & --- \\
 & TABSurfer & 0.63435±0.016 & 0.76336±0.009 & --- \\
\hline
IXI & \textbf{MedSegMamba} & \textbf{0.80858±0.035} & \textbf{0.96461±0.010} & \textbf{0.22449±0.052} \\
 & SegMambaBot & 0.77445±0.036 & 0.92566±0.010 & --- \\
 & TABSurfer & 0.62394±0.033 & 0.75991±0.009 & --- \\
\hline
NIFD & \textbf{MedSegMamba} & \textbf{0.81438±0.045} & \textbf{0.96871±0.008} & \textbf{0.22214±0.062} \\
 & SegMambaBot & 0.77286±0.051 & 0.92698±0.007 & --- \\
 & TABSurfer & 0.61505±0.043 & 0.76174±0.006 & --- \\
\hline
OAS1 & \textbf{MedSegMamba} & \textbf{0.79368±0.077} & \textbf{0.95663±0.027} & \textbf{0.27493±0.288} \\
 & SegMambaBot & 0.75796±0.072 & 0.91743±0.027 & --- \\
 & TABSurfer & 0.61299±0.061 & 0.75517±0.020 & --- \\
\hline
Overall & \textbf{MedSegMamba} & \textbf{0.80683±0.046} & \textbf{0.96299±0.015} & \textbf{0.22960±0.115} \\
 & SegMambaBot & 0.77219±0.043 & 0.92379±0.014 & --- \\
 & TABSurfer & 0.62276±0.037 & 0.75979±0.012 & --- \\
\bottomrule
\end{tabular}
\caption{Comparing MedSegMamba, SegMambaBot, and TABSurfer metrics across the hippocampal subfield segmentation datasets. Bold text indicates superior performance. ↑ indicates that higher numbers correspond to better performance and ↓ indicates that lower numbers correspond to better performance. *ASSD is not shown for SegMambaBot and TABSurfer because it cannot be computed for the classes which SegMambaBot and TABSurfer failed to learn.}
    \label{table:hsf}  
\end{table}

Metrics evaluating MedSegMamba, SegMambaBot, and TABSurfer’s hippocampal segmentations are shown in Table~\ref{table:hsf}. MedSegMamba’s improvements over both SegMambaBot and TABSurfer are more notable for this task than for subcortical segmentation, as seen in Figure~\ref{fig:parametersvsdice}b. Likely due to the smaller structures present in this task, \underline{MedSegMamba was the only network that learned to} \underline{predict all 25 classes}, while SegMambabot only learned 24 classes, and TABSurfer converged on predicting just 20 classes. This resulted in significant improvements over both models in DSC and VS. ASSD cannot be computed for missing classes, so statistical tests are not applicable between the models’ overall ASSD. However, even when ignoring the missing classes, SegMambaBot and TABSurfer get an average ASSD of  0.23380 and 0.25512, respectively, which are both still inferior to MedSegMamba’s average of 0.22960. The 3D and 2D visualizations of a sample’s hippocampus segmentation by each method are visualized in Figure~\ref{fig:3d_hsf} and Figure~\ref{fig:2d_hsf}. TABSurfer failed to segment numerous classes, leaving large regions missing. While SegMambaBot and MedSegMamba appear similar,  SegMambaBot is missing the left hippocampal amygdala transition area and slightly under-segments some regions, leaving MedSegMamba as the only model to succeed in segmenting each subregion.

\section{Discussion}
\label{sec:conclusion}

This study introduces the SS3D module and VSS3D block, an extension of Mamba's selective scan operation tailored for complex 3D image processing tasks. Integrated within the MedSegMamba hybrid CNN-Mamba architecture, the VSS3D block demonstrates robust latent-space learning capabilities, outperforming existing traditional and deep learning approaches across multiple datasets. 

The transition from a Transformer bottleneck to a Mamba-based bottleneck significantly enhances memory efficiency, thereby enabling the use of more convolutional layers for the encoder and decoder on identical hardware. This enhancement boosts local feature extraction capabilities while preserving global context extraction, resulting in both SegMambaBot and MedSegMamba outperforming TABSurfer. Comparing SegMambaBot and MedSegMamba in subcortical segmentation, the SS3D module allows MedSegMamba to attain comparable overall performance in terms of DSC and VS metrics but offers superior region boundary delineation as reflected by the ASSD, all while utilizing significantly fewer parameters than the Tri-oriented Mamba module. Increasing the number of selective scan operations to process variously unraveled sequences can achieve performance parity with larger modules that process fewer sequences. 

Compared to FastSurfer's standard 2.5D approach, our utilization of 3D inputs preserves more intricate spatial relationships within the anatomy's continuity than 2D slices. The 3D strided reconstruction process, which aggregates model outputs from shifted overlapping patches, further mitigates noise artifacts present in the ground truth. This improvement is observed qualitatively when comparing the noisier segmentations of FastSurfer to the smoother outputs of the 3D patch-based models.

The hippocampal subfield segmentation experiment also provides evidence of the SS3D module’s ability to process finer details compared to the Tri-oriented Mamba module. While SegMambaBot and TABSurfer failed this task, MedSegMamba was able to reliably segment all the small and intricate subregions.

\subsection{Limitations}
\label{ssec:limit}
The primary limitation of this approach is that MedSegMamba exhibits slower performance than SegMambaBot despite its lower parameter count, most notably during the subcortical segmentation pipeline. However, this issue can be mitigated by increasing the step size during the strided reconstruction process from 16 to 32, resulting in a roughly 4x speedup, reducing the processing time to around 22 seconds per scan with minimal impact on performance. When both use a step size of 32, MedSegMamba’s subcortical segmentation is only a few seconds slower than SegMambaBot per scan, making them roughly comparable in terms of overall efficiency while retaining the superior performance afforded by the SS3D module.
\subsection{Conclusion}
\label{ssec:conclusion}
These results showcase the advantages of this novel hybrid architecture combined with 3D patch-based processing. Future research should explore the application of our SS3D module across different architectures, including its integration at various latent space levels throughout the encoder, as seen in SegMamba's original design. Ablation studies focused on determining the optimal number of scanning directions or unraveling methods for 3D segmentation may further enhance our model’s performance.

\section{Acknowledgments}
\label{sec:acknowledgments}

No funding was received for conducting this study and there are no relevant financial or non-financial interests to disclose. 

%
%
%
\bibliographystyle{splncs04}
\bibliography{mybibliography}
%

\end{document}